\documentstyle[11pt,psfig,twoside]{article}

\begin{document}

\newcommand{\dd}{\,{\rm d}}
\newcommand{\ie}{{\it i.e.},\,}
\newcommand{\etal}{{\it et al.\ }}
\newcommand{\eg}{{\it e.g.},\,}
\newcommand{\cf}{{\it cf.\ }}
\newcommand{\vs}{{\it vs.\ }}
\newcommand{\zdot}{\makebox[0pt][l]{.}}
\newcommand{\up}[1]{\ifmmode^{\rm #1}\else$^{\rm #1}$\fi}
\newcommand{\dn}[1]{\ifmmode_{\rm #1}\else$_{\rm #1}$\fi}
\newcommand{\upd}{\up{d}}
\newcommand{\uph}{\up{h}}
\newcommand{\upm}{\up{m}}
\newcommand{\ups}{\up{s}}
\newcommand{\arcd}{\ifmmode^{\circ}\else$^{\circ}$\fi}
\newcommand{\arcm}{\ifmmode{'}\else$'$\fi}
\newcommand{\arcs}{\ifmmode{''}\else$''$\fi}
\newcommand{\MS}{{\rm M}\ifmmode_{\odot}\else$_{\odot}$\fi}
\newcommand{\RS}{{\rm R}\ifmmode_{\odot}\else$_{\odot}$\fi}
\newcommand{\LS}{{\rm L}\ifmmode_{\odot}\else$_{\odot}$\fi}

\newcommand{\Abstract}[2]{{\footnotesize\begin{center}ABSTRACT\end{center}
\vspace{1mm}\par#1\par
\noindent
{~}{\it #2}}}

\newcommand{\TabCap}[2]{\begin{center}\parbox[t]{#1}{\begin{center}
  \small {\spaceskip 2pt plus 1pt minus 1pt T a b l e}
  \refstepcounter{table}\thetable \\[2mm]
  \footnotesize #2 \end{center}}\end{center}}

\newcommand{\TableSep}[2]{\begin{table}[p]\vspace{#1}
\TabCap{#2}\end{table}}

\newcommand{\FigCap}[1]{\footnotesize\par\noindent Fig.\  %
  \refstepcounter{figure}\thefigure. #1\par}

\newcommand{\TableFont}{\footnotesize}
\newcommand{\TableFontIt}{\ttit}
\newcommand{\SetTableFont}[1]{\renewcommand{\TableFont}{#1}}

\newcommand{\MakeTable}[4]{\begin{table}[htb]\TabCap{#2}{#3}
  \begin{center} \TableFont \begin{tabular}{#1} #4 
  \end{tabular}\end{center}\end{table}}

\newcommand{\MakeTableSep}[4]{\begin{table}[p]\TabCap{#2}{#3}
  \begin{center} \TableFont \begin{tabular}{#1} #4 
  \end{tabular}\end{center}\end{table}}

\newenvironment{references}%
{
\footnotesize \frenchspacing
\renewcommand{\thesection}{}
\renewcommand{\in}{{\rm in }}
\renewcommand{\AA}{Astron.\ Astrophys.}
\newcommand{\AAS}{Astron.~Astrophys.~Suppl.~Ser.}
\newcommand{\ApJ}{Astrophys.\ J.}
\newcommand{\ApJS}{Astrophys.\ J.~Suppl.~Ser.}
\newcommand{\ApJL}{Astrophys.\ J.~Letters}
\newcommand{\AJ}{Astron.\ J.}
\newcommand{\IBVS}{IBVS}
\newcommand{\PASP}{P.A.S.P.}
\newcommand{\Acta}{Acta Astron.}
\newcommand{\MNRAS}{MNRAS}
\renewcommand{\and}{{\rm and }}
\section{{\rm REFERENCES}}
\sloppy \hyphenpenalty10000
\begin{list}{}{\leftmargin1cm\listparindent-1cm
\itemindent\listparindent\parsep0pt\itemsep0pt}}%
{\end{list}\vspace{2mm}}

\def\TYLDA{~}
\newlength{\DW}
\settowidth{\DW}{0}
\newcommand{\dw}{\hspace{\DW}}

\newcommand{\refitem}[5]{\item[]{#1} #2%
\def\REFARG{#3}\ifx\REFARG\TYLDA\else, {\it#3}\fi
\def\REFARG{#4}\ifx\REFARG\TYLDA\else, {\bf#4}\fi
\def\REFARG{#5}\ifx\REFARG\TYLDA\else, {#5}\fi.}

\newcommand{\Section}[1]{\section{#1}}
\newcommand{\Subsection}[1]{\subsection{#1}}
\newcommand{\Acknow}[1]{\par\vspace{5mm}{\bf Acknowledgments.} #1}
\pagestyle{myheadings}

\def\thefootnote{\fnsymbol{footnote}}

\begin{center}
{\Large\bf Gamma-Ray Bursts at Low Redshift}
\vskip1cm
{\bf
B~o~h~d~a~n~~P~a~c~z~y~{\'n}~s~k~i}
\vskip3mm
{Princeton University Observatory, Princeton, NJ 08544-1001, USA\\
e-mail: bp@astro.princeton.edu}
\end{center}

\Abstract{
Long duration gamma-ray bursts (GRB) are at cosmological distance, they
appear to be located near star forming regions, and are likely associated
with some type of supernovae.  They are also likely to be strongly beamed,
which lowers their energetics by several orders of magnitude, and increases
their rate by the same factor.  Therefore, it is likely that one out of
100 -- 1000 core collapse supernovae generates ultra-relativistic jets, which
beam gamma-rays and all other early emission into two narrow cones.  After a 
year, or so, the
jets are decelerated and become non-relativistic, and their emission becomes
more or less isotropic.  At least two GRB: 970508 ($ z = 0.835 $) and
980703 ($ z = 0.966$) show strong radio emission from late, and therefore
non-relativistic afterglows.  Such events should be readily detectable at low
redshift, say $ z = 0.03 $.  A search for strong radio emitters among recent
nearby supernovae should be done $ \sim 1 $ year after the explosion.
If some of these explosions generated GRB and their gamma-ray beam missed us,
the bipolar structure of the radio remnant should be resolvable with VLBA.
}{~}

\noindent
{\bf Key words:}{\it gamma-rays: bursts - ISM: supernova remnants - 
stars: supernovae}

\vskip1cm


A complete list of supernovae can be found at:\\
\centerline{http://cfa-www.harvard.edu/cfa/ps/lists/Supernovae.html}\\
The numbers discovered in the last decade, in the years 1990 -- 2000 are
38, 64, 73, 38, 40, 57, 93, 163, 161, 204, 173, respectively.  A rapid
increase is apparent, and it will continue provided the field remains
interesting.  The purpose of this paper is to point out that we may learn
about the nature of gamma-ray bursts (GRB) by studying radio remnants of
low redshift supernovae (SN).  The more SN are discovered, the better the
chance that GRB radio remnants will be identified.

The first evidence that GRB may be exploding near star forming regions
was provided by the indication of dust extinction in optical afterglows, 
and gas absorption
in X-ray afterglows (cf. Paczy\'nski 1998, Bloom et al. 1998, and references
therein).  The first direct evidence that some GRB are related to some SN
came with the SN 1998bw, which was associated with the GRB 980425 (Galama et 
al. 1998).  Less direct evidence for a GRB -- SN connection was found
for the GRB 980326 (Bloom et al. 1999) and for GRB 970228 (Reichart 1999).
While there is no proof that GRB are generated by some type of
core collapse massive supernovae, this is the most popular scenario today.
It was first suggested by Woosley (1993).

It is also popular to think that GRB emission is strongly beamed, which
reduces the total energy requirement to $ \sim 10^{51} $ erg (Frail, Waxman
and Kulkarni 2000; Frail et al. 2001; Freedman and Waxman 2001; Panaitescu
and Kumar 2001), and implies that the true GRB rate
may be 100 - 1000 times higher than observed, as we miss most of their beams.
If the GRB are indeed related to SN, this would make the true GRB rate
lower than the rate of core collapse SN by a factor 100 -- 1000.  This
makes them rare events, but note that the total number of detected supernovae
has already exceeded $ 10^3 $.  Some of them might have been related to
undetectable GRB.  In fact Cen (1999) suggested that SN 1987A might have 
been such a case.

Let us suppose that the initial GRB explosion is jet like (Paczy\'nski 1993).
The ultra-relativistic ejecta are slowed down to non-relativistic velocity
within a year or so, their emission becomes isotropic, but they retain a
highly non-spherical, bipolar geometry for many years (Ayal and
Piran 2000).  Late radio emission, originating from non-relativistic GRB
remnant was observed in at least two cases: GRB 970508 (Frail, Waxman and 
Kulkarni 2000) and GRB 980703 (Frail 2001).  It was readily detectable from
a fairly large redshift, $ z \approx 0.9 $.

Let us assume that relativistic expansion lasts 1.5 years, so a linear
extent of a remnant is $ \sim 1 $ pc.  For this to be resolved with the
VLBA its angular extent has to be larger than $ \sim 1 $ mas.  This
corresponds to the source distance $ D \sim 200 $ Mpc.  
According to Filippenko (2001, Table) the rate of core collapse SN is 0.48 per
100 yr per $ 10^{10} ~ L_{\odot}^B $.  According to Nagamine et al. (2001) 
the blue luminosity density in the local Universe is $ \sim 2.4 \times 10^8 ~
L_{\odot}^B $ per $ Mpc^3 $.  Combining these two numbers we obtain
for $ r_{SN,c} $, i.e. the rate of core collapse SN:
$$
r_{SN,c} \approx 4 \times 10^3 ~ yr^{-1} 
~ \left( { D \over 200 ~ Mpc } \right) ^3 ,
$$
while the total GRB rate is expected to be roughly:
$$
r_{GRB} \approx  10 ~ yr^{-1} ~ \left( { D \over 200 ~ Mpc } \right) ^3 .
$$
The GRB remnants are expected to remain distinctly non-spherical for 
many years (Ayal and Piran 2000), so there may be up to $ \sim 10^4 $
of them resolvable with the VLBA.  We do not know for how long they remain
bright enough for the VLBA to see, but presumably for at least a few years, as
at $ z \approx 0.05 $ their flux would be several hundred times higher that
for sources at $ z \approx 0.9 $ (GRB 970508 and GRB 980703).  While
there is no quantitative theory to support our empirical reasoning,
it seems to be likely that at any time there may be several dozen 
relatively nearby radio GRB remnants which can be resolved with VLBA as
being bipolar rather than spherical.  Note: an ordinary SN remnant would
be unresolved at $ z \approx 0.05 $, as its initial expansion is $ \sim 5,000 ~
km ~ s^{-1} $, rather than $ \sim 300,000 ~ km ~ s^{-1} $.

Over $ 10^3 $ SN were discovered since 1990, and the number is rapidly
increasing.  A fair fraction were core collapse SN, so there is a modest
chance that one or two bipolar radio GRB remnants may be
detectable among them with VLBA.
If there is none, the discovery rate of nearby SN is relatively easy
to increase, duplicating the very successful Katzman Automatic Imaging 
Telescope (KAIT, Richmond, Treffers, and Filippenko 1993):\\
\centerline{http://astron.berkeley.edu/~bait/kait.html}
which discovered almost 40 nearby SN in 2000.

A discovery of a young bipolar SN remnant would be a proof that some SN
generate bipolar relativistic ejections.  This by itself would not be a proof
that these were GRB.  However, there may be detectable imprint of a strong
gamma-ray pulse in the interstellar medium near the event (Draine 2000).
We may learn something important about GRB phenomenon studying the 
nearby Universe and using ground based instruments.

    Woods and Loeb (1999) suggested a search for $ \sim 10^3 $ years-old
GRB remnants in the Virgo cluster.  They assumed that GRB is $ \sim 10^2 -
10^3 $ times more energetic than SN, which would make an old GRB remnant 
appear distinctly different than an old SN remnant.  If GRB emission is
beamed and the total energetics is reduced to $ \sim 10^{51} $ ergs, there
is not much difference expected between old GRB and old SN remnants.  
However, the two should look very different $ \sim 1 $ year after the
explosion.

    Vietri and Stella (1999) proposed a `supranova' model for a GRB.  In
this scenario a GRB explosion happens some time after SN explosion, with
the time interval being effectively a free parameter.  Also, the initially
beamed GRB afterglow may be invisible from a random direction.  Therefore,
it is essential to look for the radio remnant and its bipolar structure
$ \sim 1 $ year after the explosion.

\Acknow{This work was not supported with any grant.}


\end{document}